\documentstyle[12pt]{article}
\def\beq{\begin{equation}}
\def\eeq{\end{equation}}
\def\bea{\begin{eqnarray}}
\def\eea{\end{eqnarray}}
\def\pr{\prime}
\newcommand{\cl}{\centerline}
\renewcommand{\theequation}{\thesection.\arbic{equation}}
\renewcommand{\theequation}{A.\arbic{equation}}
\begin{document}
\begin{titlepage}
\setlength{\textwidth}{5.0in}
\setlength{\textheight}{7.5in}
\setlength{\parskip}{0.0in}
\setlength{\baselineskip}{18.2pt}
\hfill
{\tt SOGANG-HEP 300/02}
\vskip 0.5cm
\cl{\large{\bf (2+1) dimensional black holes in warped product scheme}}\par
\vskip 1.0cm
\cl{Soon-Tae Hong$^{1,2}$, Jaedong Choi$^{3}$ and Young-Jai Park $^{2}$}
\vskip 0.4cm
\cl{$^{1}$Department of Science Education}
\cl{Ewha Womans University, Seoul 120-750, Korea}
\vskip 0.4cm
\cl{$^{2}$Department of Physics and Basic Science Research
Institute,}
\cl{Sogang University, C.P.O. Box 1142, Seoul 100-611, Korea}
\vskip 0.4cm
\cl{$^{3}$Department of Mathematics, Korea Air Force Academy}
\cl{P.O. Box 335-2, Cheongwon, Chungbuk 363-849, Korea}
\vfill
\begin{center}
{\bf ABSTRACT}
\end{center}
\begin{quote}
Exploiting a multiply warped product manifold scheme, we study the interior 
solutions of the Banados-Teitelboim-Zanelli black holes and the 
exterior solutions of the de Sitter black holes in the (2+1) dimensions. 
\end{quote}

\vskip 0.2cm
\par
\end{titlepage}

\newpage
\section{Introduction}
\setcounter{equation}{0}
\renewcommand{\theequation}{\arabic{section}.\arabic{equation}}

The Banados-Teitelboim-Zanelli (BTZ)~\cite{btz1} black hole theory 
treats special solutions of (2+1) dimensional anti-de Sitter (AdS) gravity 
possessing all the properties of black holes.  Charged BTZ black holes are 
also constructed as the analogous solutions in (2+1) dimensional AdS-Maxwell 
gravity~\cite{cal}.  Even though the BTZ black hole is a toy model in some 
respect, this BTZ black hole has triggered significant interests due to its 
connections with some string theories~\cite{string}, its role in 
microscopic entropy derivations~\cite{cal95} and quantum corrected 
thermodynamics~\cite{man97}.  Moreover, the entropy and thermodynamic 
properties of de Sitter space have been investigated~\cite{dss} to consider 
the emerging possibility that the real universe resembles de Sitter space.  
By embedding de Sitter space as a solution of string theory, various string 
dualities have been exploited to obtain a microscopic description.  However, 
so far persistent efforts have failed unfortunately even to find a completely 
satisfactory de Sitter solution of string theory.  Hopefully this situation 
will change in near future.  Recently, exploiting the global embedding Minkowski space 
approach~\cite{rosen}, several 
authors have shown that this approach could yield a unified
derivation of thermodynamics for various curved manifolds in (2+1) 
dimensions~\cite{des,mann93} 
and in (3+1) dimensions~\cite{des,andri}.  

However, all these solutions have been incompletely constructed only 
on the partial patches bounded by the event horizons of the black holes.  
On the other hand, the concept of a warped product manifold was 
introduced~\cite{bishop69} to provide a class of complete 
Riemannian manifolds with everywhere negative curvature~\cite{bo}, and was developed 
to point out that some well-known exact solutions to Einstein field equations 
are pseudo-Riemannian warped products~\cite{be}.  Furthermore, certain causal and 
completeness properties of a spacetime could be determined by 
the presence of a warped product structure~\cite{bpe}, and general theory of warped products 
were applied to discuss the special cases of Robertson-Walker and Schwarzshild manifold.  
The role of warped products in the study of exact solutions to Einstein field equations is 
now firmly established to generate interest in other areas of geometry. Recently, 
the warped product scheme has been applied to higher dimensional theories such as 
the Randall-Sundrum model~\cite{randall,rubakov,ito} in five dimension and the non-singular 
warped Kaluza-Klein embeddings~\cite{cvetic} in five to seven 
dimensional gauged supergravity theories.  Moreover, the warped product scheme was 
applied to investigate warping functions associated with constant scalar curvature 
on globally null manifold~\cite{duggal}.  Assuming the four dimensional spacetime 
to be a warped product of two surfaces, the four dimensional Einstein equations were 
also reduced to two dimensional ones to describe wormwholes and domainwalls of 
curvature singularities~\cite{kata}.   

In order to investigate physical properties inside the black hole horizons, we
briefly review a multiply warped product manifold $(M=B\times F_1\times...\times F_{n}, g)$ 
which consists of the Riemannian base manifold 
$(B, g_B)$ and fibers $(F_i,g_i)$ ($i=1,...,n$) associated with the Lorentzian metric,
\beq
g=\pi_{B}^{*}g_{B}+\sum_{i=1}^{n}(f_{i}\circ\pi_{B})^{2}\pi_{i}^{*}g_{i}
\label{g}
\eeq
where $\pi_B$, $\pi_{i}$ are the natural projections of $B\times F_1\times...\times F_n$ 
onto $B$ and $F_{i}$, respectively, and $f_{i}$ are positive warping functions.  For the specific 
case of $(B=R,g_B=-d\mu^{2})$, the above metric is rewritten as 
\beq
g=-d\mu^{2}+\sum_{i=1}^{n}f_{i}^{2}g_{i},
\label{gnew}
\eeq
to extend the warped product spaces to richer class of spaces involving multiply 
products. Moreover, the conditions of spacelike boundaries in the multiply warped 
product spacetimes~\cite{fs} were also studied~\cite{ha} and the curvature of the 
multiply warped product with $C^0$-warping functions was later investigated~\cite{choi00}.  
From a physical point of view, these warped product spacetimes 
are interesting since they include classical examples of spacetime such as the Robertson-Walker 
manifold and the intermediate zone of RN manifold~\cite{rn,ksy}.  Recently, 
the interior Schwarzschild spacetime has been represented as a multiply warped
product spacetime with warping functions~\cite{choi00} to yield the Ricci curvature 
in terms of $f_1$ and $f_2$ for the multiply warped products of the form 
$M=R\times_{f_1}R\times_{f_2} S^2$. Very recently, we have studied the interior RN-AdS spacetime
by exploiting this multiply warped product scheme~\cite{hcp02}.

In this paper we will analyze the multiply warped product manifold associated 
with the charge black holes such as the BTZ and de
Sitter (dS) metrics to investigate the physical
properties inside the event horizons.  We will exploit the multiply warped product scheme 
to investigate the interior solutions in (2+1) charged BTZ black holes in section 2, in 
(2+1) charged dS black holes in section 3 so that we can explicitly obtain the Ricci 
and Einstein curvatures inside the event horizons of these metrics.

\section{BTZ black holes }
\setcounter{equation}{0}
\renewcommand{\theequation}{\arabic{section}.\arabic{equation}}
\subsection{Static BTZ case}

In order to investigate a multiply warped product manifold for
the static  Banados-Teitelboim-Zanelli (BTZ) interior solution, we start 
with the three-metric inside the horizon   
\beq
ds^{2}=N^{2}dt^{2}-N^{-2}dr^{2}+r^{2}d\phi^{2}
\label{btzmetric}
\eeq
with the lapse function for the interior solution
\beq
N^{2}=m-\frac{r^{2}}{l^{2}}.
\label{btzlapse}
\eeq
Note that the event horizon $r_{H}$ is given by $r_{H}=m^{1/2}l$.
Furthermore the lapse function can be rewritten in terms of the event horizon 
as follows
\beq
N^{2}=\frac{(r_{H}+r)(r_{H}-r)}{l^{2}}
\label{lapserh}
\eeq
which is well defined in the region $r<r_{H}$.  

Now we define a new coordinate $\mu$ as follows
\beq
d\mu^{2}=N^{-2}dr^{2},
\label{btzdmu}
\eeq
which can be integrated to yield
\beq
\mu=\int_{0}^{r}dx~\frac{l}{[(r_{H}+x)(r_{H}-x)]^{1/2}},
\eeq
whose analytic solution is of the form
\beq
\mu=l\sin^{-1}\left(\frac{r}{r_{H}}\right)=F(r).
\label{btzsolmu}
\eeq
Moreover, we have the following boundary conditions
\beq
{\rm lim}_{r\rightarrow r_{H}}F(r)=\frac{l\pi}{2},~~~
{\rm lim}_{r\rightarrow 0}F(r)=0,
\label{btzbdy}
\eeq
and $dr/d\mu >0$ implies $F^{-1}$ is well-defined function.
 
Exploiting the above new coordinate (\ref{btzsolmu}), we rewrite the
metric (\ref{btzmetric}) 
as a warped products 
\beq
ds^{2}=-d\mu^{2}+f_{1}(\mu)^{2}dt^{2}+f_{2}^{2}(\mu)d\phi^{2}
\label{btzmetric2}
\eeq
where 
\bea
f_{1}(\mu)&=&\left(m-\frac{F^{-2}(\mu)}{l^{2}}\right)^{1/2},
\nonumber\\
f_{2}(\mu)&=&F^{-1}(\mu).
\label{btzf1f2}
\eea

After some algebra, we obtain the following nonvanishing Ricci curvature components
\bea
R_{\mu\mu}&=&-\frac{f_{1}^{\pr\pr}}{f_{1}}-\frac{f_{2}^{\pr\pr}}{f_{2}},
\nonumber\\
R_{tt}&=&\frac{f_{1}f_{1}^{\pr}f_{2}^{\pr}}{f_{2}}+f_{1}f_{1}^{\pr\pr},
\nonumber\\
R_{\phi\phi}&=&\frac{f_{1}^{\pr}f_{2}f_{2}^{\pr}}{f_{1}}+f_{2}f_{2}^{\pr\pr}.
\label{btzricci}
\eea

Using the explicit expressions for $f_{1}$ and $f_{2}$ in (\ref{btzf1f2}), one can obtain 
identities for $f_{1}$, $f_{1}^{\pr}$ and $f_{1}^{\pr\pr}$ in terms of $f_{1}$, $f_{2}$ 
and their derivatives
\bea
f_{1}&=&f_{2}^{\pr},\nonumber\\
f_{1}^{\pr}&=&-\frac{f_{2}}{l^{2}},\nonumber\\
f_{1}^{\pr\pr}&=&\frac{f_{1}f_{1}^{\pr}}{f_{2}},
\label{btzids}
\eea  
to yield the Ricci curvature components
\bea
R_{\mu\mu}&=&-\frac{2f_{1}^{\pr}}{f_{2}},\nonumber\\ 
R_{tt}&=&\frac{2f_{1}^{2}f_{1}^{\pr}}{f_{2}},\nonumber\\
R_{\phi\phi}&=&2f_{2}f_{1}^{\pr},
\label{btzriccis}
\eea
and the Einstein scalar curvature 
\beq
R=-\frac{6}{l^{2}},
\label{btzeinr}
\eeq
in the interior of the static BTZ black hole horizon.

\subsection{Charged BTZ case}

Now we consider a multiply warped product manifold associated with the
charged BTZ three-metric (\ref{btzmetric}) inside the horizon with the 
charged lapse function~\cite{cal} 
\beq
N^{2}=m-\frac{r^{2}}{l^{2}}+2Q^{2}\ln~r.
\label{cbtzlapse}
\eeq
Note that the event horizon $r_{H}$ satisfies the equation 
$0=m-\frac{r_{H}^{2}}{l^{2}}+2Q^{2}\ln~r_{H}$, and for the range $Ql<r<r_{H}$
we have the coordinate $\mu$ in Eq. (\ref{btzdmu}) 
\beq
\mu=\int_{Ql}^{r}dx~\frac{l}{(m-\frac{r^{2}}{l^{2}}+2Q^{2}\ln~r)^{1/2}},
\label{cbtzmu}
\eeq
so that $dr/d\mu >0$ implies $F^{-1}$ is well-defined function.
 
Exploiting the above coordinate (\ref{cbtzmu}), we can obtain 
the warped products (\ref{btzmetric2}) with the modified $f_{1}$ and
$f_{2}$ as below
\bea
f_{1}(\mu)&=&\left(m-\frac{F^{-2}(\mu)}{l^{2}}+2Q^{2}\ln~F^{-1}(\mu)\right)^{1/2},
\nonumber\\
f_{2}(\mu)&=&F^{-1}(\mu),
\label{cbtzf1f2}
\eea
to yield the Ricci curvature components
\bea
R_{\mu\mu}&=&-\frac{2f_{1}^{\pr}}{f_{2}}+\frac{2Q^{2}}{f_{2}^{2}},\nonumber\\ 
R_{tt}&=&\frac{2f_{1}^{2}f_{1}^{\pr}}{f_{2}}-\frac{2Q^{2}f_{1}^{2}}{f_{2}^{2}},\nonumber\\
R_{\phi\phi}&=&2f_{2}f_{1}^{\pr},
\label{cbtzriccis}
\eea
and the Einstein scalar curvature
\beq
R=-\frac{6}{l^{2}}+\frac{2Q^{2}}{f_{2}^{2}},
\label{cbtzeinr}
\eeq
in the interior of the charged BTZ black hole horizon.  
 
Now it seems appropriate to comment on the relations between the interior and exterior solutions 
in the charged BTZ black hole.  In the exterior of the event horizon $r_{H}$ where the three-metric is 
given by
\beq
ds^{2}=-(-m+\frac{r^{2}}{l^{2}}-2Q^{2}\ln~r)^{2}dt^{2}
+(-m+\frac{r^{2}}{l^{2}}-2Q^{2}\ln~r)^{-2}dr^{2}+r^{2}d\phi^{2},
\label{nsmetric3}
\eeq
one can obtain the Ricci curvature components in terms of the warping functions $f_{1}$ and 
$f_{2}$ as follows 
\bea
R_{rr}&=&-\frac{2f_{1}^{\pr}}{f_{1}^{2}f_{2}}+\frac{2Q^{2}}{f_{1}^{2}f_{2}^{2}},\nonumber\\ 
R_{tt}&=&\frac{2f_{1}^{2}f_{1}^{\pr}}{f_{2}}-\frac{2Q^{2}f_{1}^{2}}{f_{2}^{2}},\nonumber\\
R_{\phi\phi}&=&2f_{2}f_{1}^{\pr},
\label{adsricci2}
\eea
and the Einstein scalar curvature identical to the interior case (\ref{cbtzeinr}).  Here one notes that 
the Ricci components $R_{tt}$ and $R_{\phi\phi}$ are the same as those of interior case.  
Moreover from the definition of the coordinate $\mu$ in Eq. (\ref{btzdmu}) one can obtain the identity 
\beq
R_{\mu\mu}=f_{1}^{2}R_{rr}
\eeq
which is also attainable from the Ricci components $R_{\mu\mu}$ and $R_{rr}$ in Eqs. (\ref{cbtzriccis}) 
and (\ref{adsricci2}).  One can thus show that all the Ricci components and the  Einstein scalar curvature 
are identical both in the exterior and interior of the event horizon $r_{H}$ without discontinuities.

\subsection{Rotating BTZ case}

Now we consider a multiply warped product manifold associated with the
rotating BTZ black hole inside the horizon whose three-metric is given by 
\beq
ds^{2}=N^{2}dt^{2}-N^{-2}dr^{2}+r^{2}(d\phi+N^{\phi}dt)^{2}
\label{rotbtzmetric}
\eeq
where the lapse and shift functions are given by
\bea
N^{2}&=&m-\frac{r^{2}}{l^{2}}-\frac{J^{2}}{4r^{2}},
\nonumber\\
N^{\phi}&=&-\frac{J}{2r^{2}},
\label{rotbtzlapse}
\eea
with an angular momentum $J$.  Note that the event horizon $r_{\pm}$ satisfies the equation 
$0=m-\frac{r_{\pm}^{2}}{l^{2}}-\frac{J^{2}}{4r_{\pm}^{2}}$ to yield 
the lapse function in terms of the event horizons 
as follows
\beq
N^{2}=\frac{(r_{+}^{2}-r^{2})(r^{2}-r_{-}^{2})}{r^{2}l^{2}}
\label{rotlapserh}
\eeq
which, for the interior solution, is well defined in the region $r_{-}<r<r_{+}$.  

Defining a new coordinate $\mu$ as in Eq. (\ref{btzdmu}), we obtain 
\beq
\mu=\int_{r_{-}}^{r}dx~\frac{l}{(m-\frac{r^{2}}{l^{2}}-\frac{J^{2}}{4r^{2}})^{1/2}},
\label{rotbtzsolmu}
\eeq
whose analytic solution is of the form
\beq
\mu=l\sin^{-1}\left(\frac{r^{2}-r_{-}^{2}}{r_{+}^{2}-r_{-}^{2}}\right)^{1/2}=F(r).
\label{rotbtzan}
\eeq
Moreover, we have the following boundary conditions
\beq
{\rm lim}_{r\rightarrow r_{+}}F(r)=\frac{l\pi}{2},~~~
{\rm lim}_{r\rightarrow r_{-}}F(r)=0.
\label{rotbtzbdy}
\eeq
Note that $dr/d\mu >0$ implies $F^{-1}$ is well-defined function and in the vanishing angular 
momentum limit $J\rightarrow 0$, the above solution (\ref{rotbtzan}) reduces to the static BTZ 
case (\ref{btzsolmu}).
 
Exploiting the above new coordinate (\ref{rotbtzan}), we can obtain 
\beq
ds^{2}=-d\mu^{2}+f_{1}(\mu)^{2}dt^{2}+f_{2}^{2}(\mu)(d\phi+N^{\phi}dt)^{2}
\label{rotmetric}
\eeq
to yield the metric of the warped product form (\ref{btzmetric2}) in a comoving coordinates where 
one can replace\footnote{Here one notes that the detector 
locates in the comoving coordinates with the angular velocity 
$d\phi/dt=-g_{t\phi}/g_{\phi\phi}=-N^{\phi}$.}  $d\phi+N^{\phi}dt\rightarrow d\phi$
to obtain the modified $f_{1}$ and $f_{2}$ as below
\bea
f_{1}(\mu)&=&\left(m-\frac{F^{-2}(\mu)}{l^{2}}-\frac{J^{2}}{4F^{-2}(\mu)}\right)^{1/2},
\nonumber\\
f_{2}(\mu)&=&F^{-1}(\mu),
\label{rotbtzf1f2}
\eea
and the Ricci curvature components
\bea
R_{\mu\mu}&=&-\frac{2f_{1}^{\pr}}{f_{2}}+\frac{J^{2}}{f_{2}^{4}},\nonumber\\ 
R_{tt}&=&\frac{2f_{1}^{2}f_{1}^{\pr}}{f_{2}}-\frac{J^{2}f_{1}^{2}}{f_{2}^{4}},\nonumber\\
R_{\phi\phi}&=&2f_{2}f_{1}^{\pr}.
\label{rotbtzriccis}
\eea
Here one notes that there does not exist an additional term associated with the angular 
momentum $J$ in the $R_{\phi\phi}$ component since we have used the comoving coordinates.  
The Einstein scalar curvature is then given by 
\beq
R=-\frac{6}{l^{2}}-\frac{J^{2}}{2f_{2}^{4}},
\label{rotein}
\eeq
in the interior of the charged BTZ black hole horizons.  Note that in the $J\rightarrow 0$ limit, 
the above Ricci components (\ref{rotbtzriccis}) and Einstein scalar curvature (\ref{rotein}) 
reduce to the corresponding ones in the static BTZ case.

\section{dS black holes }
\setcounter{equation}{0}
\renewcommand{\theequation}{\arabic{section}.\arabic{equation}}
\subsection{Static dS case}

In order to investigate a multiply warped product manifold for
the static de Sitter (dS) exterior solution, we start with the three-metric 
(\ref{btzmetric}) outside the horizon with the lapse function for the 
exterior solution
\beq
N^{2}=-m+\frac{r^{2}}{l^{2}}.
\label{dslapse}
\eeq
Note the event horizon $r_{H}$ is given by $r_{H}=m^{1/2}l$.
Furthermore the lapse function can be rewritten in terms of the event horizon 
as follows
\beq
N^{2}=\frac{(r+r_{H})(r-r_{H})}{l^{2}}
\label{dslapserh}
\eeq
which is well defined in the region $r>r_{H}$.  

With a new coordinate $\mu$ as in the BTZ case we obtain 
\beq
\mu=l\cosh^{-1}\left(\frac{r}{r_{H}}\right)=F(r),
\label{dssolmu}
\eeq
and the boundary condition
\beq
{\rm lim}_{r\rightarrow r_{H}}F(r)=0.
\label{dsbdy}
\eeq
Note that $dr/d\mu >0$ implies $F^{-1}$ is well-defined function.
 
Exploiting the above new coordinate (\ref{dssolmu}), we rewrite the
metric (\ref{btzmetric}) with the lapse function (\ref{dslapse}) as a warped products 
(\ref{btzmetric2}) where 
\bea
f_{1}(\mu)&=&\left(-m+\frac{F^{-2}(\mu)}{l^{2}}\right)^{1/2},
\nonumber\\
f_{2}(\mu)&=&F^{-1}(\mu),
\label{dsf1f2}
\eea
to yield, in the exterior of the static dS black hole horizon, the
same form of Ricci curvature components (\ref{btzriccis}) as those of the
static BTZ case, and the Einstein scalar curvature
\beq
R=\frac{6}{l^{2}},
\label{dsinr}
\eeq
which has the opposite sign of the static BTZ result (\ref{btzeinr}). 
 
\subsection{Charged dS case}

Now we consider a multiply warped product manifold associated with the
charged dS three-metric (\ref{btzmetric}) outside the horizon with the 
charged lapse function 
\beq
N^{2}=-m+\frac{r^{2}}{l^{2}}+2Q^{2}\ln~r.
\label{cdslapse}
\eeq
The event horizon $r_{H}$ then satisfies the equation 
$0=-m+\frac{r_{H}^{2}}{l^{2}}+2Q^{2}\ln~r_{H}$, and for the range $r>r_{H}$
the coordinate $\mu$ in Eq. (\ref{btzdmu}) is given by 
\beq
\mu=\int_{r_{H}}^{r}dx~\frac{l}{(-m+\frac{r^{2}}{l^{2}}+2Q^{2}\ln~r)^{1/2}}.
\label{cdsmu}
\eeq
Note that $dr/d\mu >0$ implies $F^{-1}$ is well-defined function.
 
Exploiting the above coordinate (\ref{cbtzmu}), we can obtain 
the warped products (\ref{btzmetric2}) with the modified $f_{1}$ and
$f_{2}$ as below
\bea
f_{1}(\mu)&=&\left(-m+\frac{F^{-2}(\mu)}{l^{2}}+2Q^{2}\ln~F^{-1}(\mu)\right)^{1/2},
\nonumber\\
f_{2}(\mu)&=&F^{-1}(\mu),
\label{cdsf1f2}
\eea
to yield, in the exterior of the charged dS black hole horizon, the
same form of Ricci curvature components (\ref{cbtzriccis}) as those of the
charged BTZ case, and the Einstein scalar curvature
\beq
R=\frac{6}{l^{2}}+\frac{2Q^{2}}{f_{2}^{2}}.
\label{cdsinr}
\eeq

\subsection{Rotating dS case}

Now we consider a multiply warped product manifold associated with the
rotating dS black hole outside the horizon where three-metric (\ref{rotbtzmetric}) 
is given by the lapse and shift functions are now given by
\bea
N^{2}&=&-m+\frac{r^{2}}{l^{2}}-\frac{J^{2}}{4r^{2}},
\nonumber\\
N^{\phi}&=&-\frac{J}{2r^{2}}.
\label{rotdslapse}
\eea
Note that the event horizon $r_{\pm}$ satisfies the equation 
$0=-m+\frac{r_{\pm}^{2}}{l^{2}}-\frac{J^{2}}{4r_{\pm}^{2}}$ to yield 
the lapse function in terms of the event horizons 
as follows
\beq
N^{2}=\frac{(r^{2}-r_{+}^{2})(r^{2}-r_{-}^{2})}{r^{2}l^{2}},
\label{rotdslapserh}
\eeq
which, for the exterior solution, is well defined in the region $r>r_{+}$.  

With a new coordinate $\mu$ as in Eq. (\ref{btzdmu}) we obtain 
\beq
\mu=\int_{r_{+}}^{r}dx~\frac{l}{(-m+\frac{r^{2}}{l^{2}}-\frac{J^{2}}{4r^{2}})^{1/2}},
\label{rotdssolmu}
\eeq
whose analytic solution is of the form
\beq
\mu=l\cosh^{-1}\left(\frac{r^{2}-r_{-}^{2}}{r_{+}^{2}-r_{-}^{2}}\right)^{1/2}=F(r).
\label{rotdsan}
\eeq
Moreover, we have the following boundary conditions
\beq
{\rm lim}_{r\rightarrow r_{+}}F(r)=0.
\label{rotdsbdy}
\eeq
Note that $dr/d\mu >0$ implies $F^{-1}$ is well-defined function and in the vanishing angular 
momentum limit $J\rightarrow 0$, the above solution (\ref{rotdsan}) reduces to 
the static dS case (\ref{dssolmu}).
 
Exploiting the above new coordinate (\ref{rotdsan}), we can obtain the metric (\ref{rotmetric})
to yield the warped products (\ref{btzmetric2}) in a comoving coordinates where 
one can replace $d\phi+N^{\phi}dt\rightarrow d\phi$ and the modified $f_{1}$ and
$f_{2}$ are given as below
\bea
f_{1}(\mu)&=&\left(-m+\frac{F^{-2}(\mu)}{l^{2}}-\frac{J^{2}}{4F^{-2}(\mu)}\right)^{1/2},
\nonumber\\
f_{2}(\mu)&=&F^{-1}(\mu),
\label{rotdsf1f2}
\eea
to yield, in the exterior of the rotating dS black hole horizon, the
same Ricci curvature components (\ref{rotbtzriccis}) as those of the
rotating BTZ case, and the Einstein scalar curvature
\beq
R=\frac{6}{l^{2}}-\frac{J^{2}}{2f_{2}^{4}}.
\label{rotdsinr}
\eeq
Note that in the $J\rightarrow 0$ limit, the above Einstein scalar curvature (\ref{dsinr}) 
reduce to the corresponding ones in the static dS case.

\section{Conclusions}
\setcounter{equation}{0}
\renewcommand{\theequation}{\arabic{section}.\arabic{equation}}
\label{sec:conclusions}

We have studied  a multiply warped product manifold associated with the BTZ (dS) 
black holes to evaluate the Ricci curvature components 
inside (outside) the black hole horizons.  Moreover, we have shown that all the Ricci 
components and the Einstein scalar curvatures are identical both in the exterior and 
interior of the event horizons without discontinuities for both the BTZ and 
dS black holes.  Through further investigation, it will be interesting to study the 
thermodynamics for interior (exterior) solutions of BTZ (dS) black holes. 

\vskip 0.5cm

STH and JC would like to acknowledge financial support in part
from Korea Science and Engineering Foundation Grant (R01-2000-00015) and 
(R01-2001-000-00003-0), respectively.


\begin{thebibliography}{99}
\bibitem{btz1} M. Banados, C. Teitelboim and J. Zanelli, Phys. Rev. Let.
{\bf 69}, 1849 (1992); M. Banados, M. Henneaux, C. Teitelboim and
J. Zanelli, Phys. Rev. {\bf D48}, 1506 (1993).
\bibitem{cal} G. Clement, Class. Quant. Grav. {\bf 10}, L49 (1993);  
S. Carlip, Class. Quant. Grav. {\bf 12}, 2853 (1995); 
G. Clement, Phys. Lett. {\bf B367}, 70(1996), C. Martinez, C. Teitelboim 
and J. Zanelli, Phys. Rev. {\bf D61}, 104013 (2000).
\bibitem{string} S. Hyun, {\tt hep-th/9704005} (1997); K. Sfetsos and K. Skenderis, 
Nucl. Phys. {\bf B517}, 179 (1998).
\bibitem{cal95} S. Carlip, Phys. Rev. {\bf D51}, 632 (1995); Phys. Rev. {\bf D55}, 878 
(1997).
\bibitem{man97} R.B. Mann and S.N. Solodukhin, Phys. Rev. {\bf D55}, 3622 (1997); 
A.J.M. Medved and G. Kunstatter, Phys. Rev. {\bf D63}, 104005 (2001); M. Buric, M. 
Dimitrijevic and V. Radovanovic, {\tt hep-th/0108036}. 
\bibitem{dss} J. Maldacena and A. Strominger, JHEP {\bf 9802}, 014 (1998); A. Strominger, 
JHEP {\bf 9802}, 009 (1998); R. Bousso, JHEP {\bf 9906}, 028 (1999);
W.T. Kim, Phys. Rev. D {\bf 59}, 047503 (1999); V. Balasubramanian, P. Horava and D. 
Minic, JHEP {\bf 0105}, 043 (2001); S. Hawking, J. Maldacena and A. Strominger, JHEP 
{\bf 0105}, 001 (2001); R. Bousso, JHEP {\bf 0104}, 035 (2001).
\bibitem{rosen} J. Rosen, Rev. Mod. Phys. {\bf 37}, 204 (1965).
\bibitem{des} S. Deser and O. Levin, Class. Quantum Grav. {\bf 14},
L163 (1997); Class. Quantum Grav. {\bf 15}, L85 (1998); Phys. Rev. {\bf D59},
0640004 (1999).
\bibitem{mann93} D. Cangemi, M. Leblanc and R.B. Mann, Phys. Rev. {\bf D48},
3606 (1993); S.T. Hong, Y.W. Kim and Y.J. Park, Phys. 
Rev. {\bf D62}, 024024 (2000); S.T. Hong, W.T. Kim, Y.W. Kim and Y.J. Park, Phys. 
Rev. {\bf D62}, 064021 (2000).
\bibitem{andri} L. Andrianopoli, M. Derix, G.W. Gibbons, C. Herdeiro, A. Santambrogio and
A. V. Proeyen, Class. Quant. Grav. {\bf 17}, 1875 (2000); S.W. Hawking and H.S. 
Reall, Phys. Rev. {\bf D61}, 024014 (1999); Y.W. Kim, Y.J. Park and K.S. Soh, 
Phys. Rev. {\bf D62}, 104020 (2000).
\bibitem{bishop69} R.L. Bishop and B. O'Neill, Am. Math. Soc. {\bf 145}, 1 
(1969).
\bibitem{bo} R.L. Bishop and B. O'Neill, Trans. A.M.S. {\bf 145}, 1 (1969).
\bibitem{be} J.K. Beem, P. E. Ehrlich and K. Easley,
{\it Global Lorentzian Geometry} (Marcel Dekker Pure and Applied Mathematics, 
New York, 1996).
\bibitem{bpe} J.K. Beem and P. E. Ehrlich, Math. Proc. Camb. Phil. Soc.
 {\bf 85}, 161 (1979).
\bibitem{randall} L. Randall and R. Sundrum, Phys. Rev. Lett. {\bf 83}, 3370 (1999);
Phys. Rev. Lett. {\bf 83}, 4690 (1999).
\bibitem{rubakov} V.A. Rubakov and M.E. Shaposhnikov, Phys. Lett. B {\bf 125}, 139 (1983).
\bibitem{ito} M. Ito, Phys. Lett. B {\bf 524}, 357 (2002); {\tt hep-th/0105186}.
\bibitem{cvetic} M. Cvetic, H. Lu and C.N. Pope, Nucl. Phys. B {\bf 597}, 172 (2001).
\bibitem{duggal} K.L. Duggal, Nonlin. Anal. {\bf 47}, 3061 (2001); J. Geom, Phys. 
{\bf 809}, in press.
\bibitem{kata} M.O. Katanaev, T. Klosch and W. Kummer, Ann. Phys. {\bf 276}, 191 (1999).
\bibitem{fs} J.L. Flores and M. S\'{a}nchez, {\tt math.DG/9909075}; {\tt math.DG/0106174}.  
\bibitem{ha} S.G. Harris, Class. Quant. Grav. {\bf 17}, 551 (2000).
\bibitem{choi00} J. Choi, J. Math. Phys. {\bf 41}, 8163 (2000).
\bibitem{rn} H. Reissner, Ann. Phys. {\bf 50}, 106 (1916);
G. Nordstr\"om, Proc. K. Ned. Akda. Wet {\bf 20}, 1238 (1918).
\bibitem{ksy} J. Demers, R. Lafrance, and R. C. Meyers, Phys. Rev. D {\bf 52},
2245 (1995); A. Ghosh and P. Mitra, Phys. Lett. B {\bf 357}, 295 (1995);
S. P. Kim, S. K. Kim, K. S. Soh, and J. H. Yee, Int. J. Mod. Phys. A {\bf 12},
5223 (1997); G. Cognola and P. Lecca, Phys. Rev. D {\bf 57}, 1108 (1998).
\bibitem{hcp02} S.T. Hong, J. Choi and Y.J. Park, ``Warped products and 
Reissner-Nordstr\"{o}m-AdS black hole".
\end{thebibliography}
\end{document}